Highlights

- A multidimensional approach evidence different level of inequalities distribution
- Groups with diverse socioeconomic condition of accessibility to services are mapped
- Spatial inequalities of a global south megacity São Paulo (Brazil) are explored
- Low income groups have low accessibility mainly to hospitals and culture centers
- A heterogeneous condition of accessibility and income in the periphery is captured

Graphical abstract

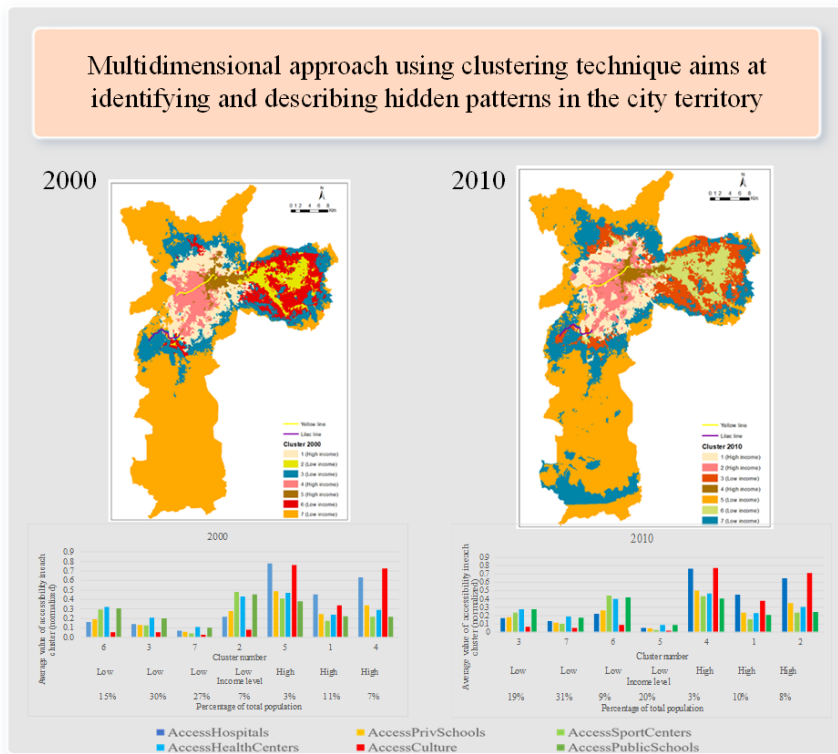

# LEARNING ABOUT SPATIAL INEQUALITIES: CAPTURING THE HETEROGENEITY IN THE URBAN ENVIRONMENT

**ABSTRACT**

*Transportation systems can be conceptualized as an instrument of spreading people and resources over the territory, playing an important role in developing sustainable cities. The current rationale of transport provision is based on population demand, disregarding land use and socioeconomic information. To meet the challenge to promote a more equitable resource distribution, this work aims at identifying and describing patterns of urban services supply, their accessibility, and household income. By using a multidimensional approach, the spatial inequalities of a large city of the global south reveal that the low-income population has low access mainly to hospitals and cultural centers. A low-income group presents an intermediate level of accessibility to public schools and sports centers, evidencing the diverse condition of citizens in the peripheries. These complex outcomes generated by the interaction of land use and public transportation emphasize the importance of comprehensive methodological approaches to support decisions of urban projects, plans and programs. Reducing spatial inequalities, especially providing services for deprived groups, is fundamental to promote the sustainable use of resources and optimize the daily commuting.*

## 1. INTRODUCTION

(In)equality is related to the distribution of opportunities, such as goods, services, infrastructure and income (Poyart et al., 2018). The territory is the ground where the distribution of these opportunities takes place, evidencing the important role of spatially explicit initiatives, instruments and policies to reduce these disparities (Fisk, 2012). With the complexity of modern cities, the infrastructure systems require adequate long-term visions, informed by Sustainable Development Goals (UN, 2015), adequate plans (Tracker et al., 2019) and decision-support tools to minimize internal transportation needs, optimize locational opportunities as well as people and material flows (Fisk, 2012).

In recent years, the role played by transport provision in distributing opportunities and people across the territory created a paradigm shift in transportation planning (Chen et al., 2018; Lucas, 2012). This rationality raised awareness for transportation plans to integrate land use plans, promoting the sustainable distribution of people and resources (Sebathu et al., 2017). However, there is a missing link between transportation planning theory and practice (Boisjoly and El-Geneidy, 2017a; Manaugh et al., 2015) and the integration of environmental and social issues into transportation projects, plans and programs is still overlooked in some regions, prevailing political and economic interests (Malvestio et al., 2018).

Accessibility is a suitable concept to frame this integration (Bertolini et al., 2005). It can be defined as the potential opportunities for interaction (Hansen, 1959). The literature points out a considerable number of definitions and components (Geurs & van Wee, 2004); reviews of indicators (Páez et al., 2012; van Wee, 2016); investigation about the changes in the accessibility level over the years (Foth et al., 2013); methods to explore different accessibility levels (Wang, 2018; Xing et al., 2018); identification of the current consideration related to objectives and measures in transportation plans (Boisjoly & El-Geneidy, 2017a) and how this is perceived by practitioners (Boisjoly & El-Geneidy, 2017b).

By connecting people and resources, accessibility measures aim at providing a measure of their equitable and sustainable distribution across the territory. They are developed to inform mainly the decision makers about the number, quality and availability of spatially distributed potential opportunities to be reached given a cost of travel. Although some studies advances in constructing equity measures (Neutens et al, 2010), using the Gini Index and Lorenz curve to evaluate the cumulative percentage of access regarding a specific group (Delbosc & Currie, 2011; Lucas et al., 2015), Lucas (2012) calls attention for the need of more comprehensive and innovative approaches for capturing the complexity of opportunities distribution.

For reducing spatial inequalities, the transportation plans should be supported by the information about the heterogeneity of socioeconomic, accessibility and opportunities. To address this demand, methodologies based on grouping instances in a high dimensional dataset are useful to identify different levels of services provision and socioeconomic condition of inhabitants. Dimensionality reduction is used to identify relevant features in the dataset, revealing its structure. It reduces computational costs, removes noise and makes the dataset easier to use (Harrington, 2016). Complementary to this, clustering aims at identifying and describing hidden patterns in the dataset, dividing the instances into similar groups (Harrington, 2016). To determine the similarity between groups, it considers the general distances between the features. Both techniques, if associated, identify the relevant features and cluster them into similar groups for capturing the distribution of services and infrastructure across the territory (Ibes, 2015). This multidimensional approach classifies similar areas and can be adapted to and replicated in different contexts.

Based on the motivation for capturing the heterogeneity in inequalities of opportunities distribution, this work aims at identifying and describing patterns hidden in accessibility indicators to different urban services and socioeconomic information using a multidimensional approach. The São Paulo municipality and its metropolitan region presents a distinguished condition of transport infrastructure and social groups (Boisjoly et al., 2017; Moreno-Monroy et al., 2017). Investigations focused on identifying different aspects of social issues across the São Paulo territory describing different: (i) socioeconomic distribution (Marques, 2005); (ii) levels of segregation (Feitosa et al., 2007); (iii) degree of social exclusion (Sposati; Monteiro; 2017). However, few explore the levels of accessibility to different urban services and their relationship with household income.

A clustering technique is applied to a large city of the global south, based transportation and census data, for answering the following question: how spatial inequalities are distributed considering transportation, land use and socioeconomic information? This application aims at promoting a sustainable distribution of resources, providing transport also for deprived groups . This work is structured as follows. Section two present the dataset and methods used for exploring accessibility and income. Section three presents the inequalities analysis applied to the São Paulo municipality. Section four presents the discussion of the results with the related literature and finally, the conclusions are drawn.

## 2. MATERIALS AND METHODS

For the study, we used census information from 2000 and 2010. In addition, geographic information about the transportation lines and major services (e.g. hospitals, schools and leisure facilities) were used. To answer the research question, the main steps of the investigation are summarized in Figure 1. First, the indicators of income and accessibility were calculated. The data were then analyzed using dimensionality reduction and clustering techniques. The ArcMap 10.5 was used to calculate the indicators and the maps design. The software used to perform the clustering and dimensionality reduction was Weka (Frank et al., 2016).

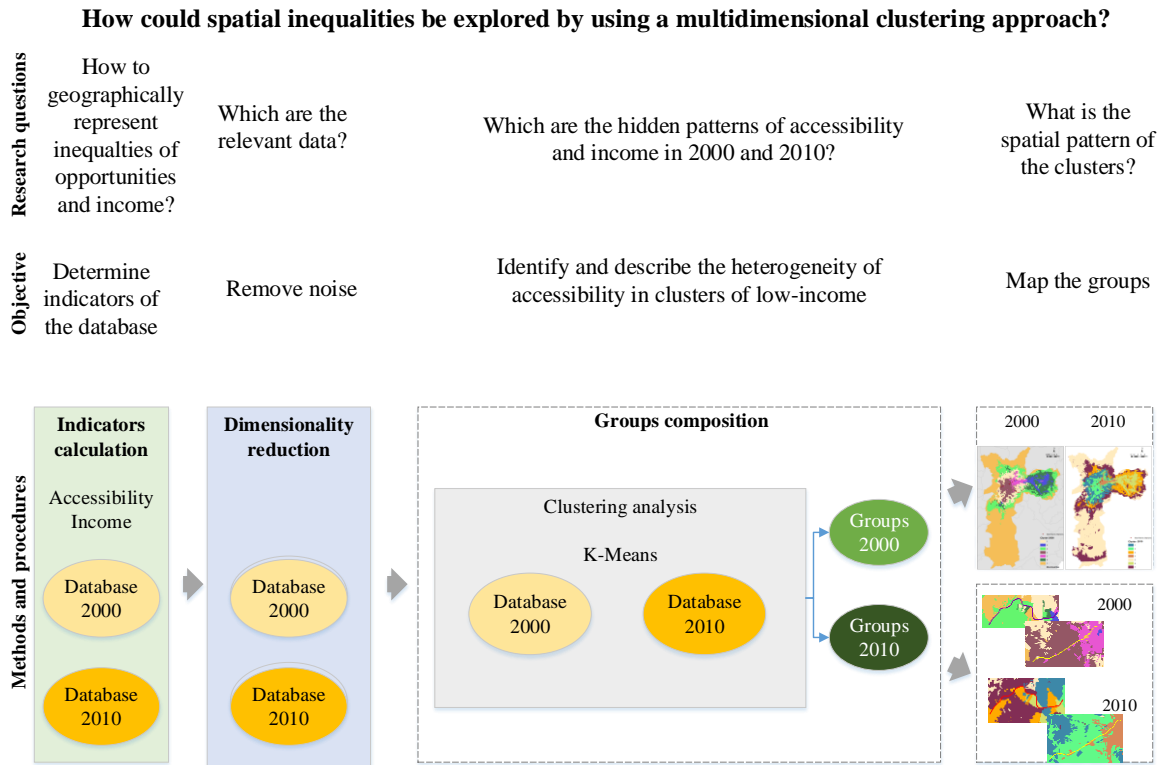

Figure 1 – Research framework of unsupervised analysis

### 2.1. Indicators calculation

Based on the configurations of metro network in the two-time periods, 2000 and 2010, the inference of transit travel time was used to estimate the travel cost for the accessibility indicators. The network was built firstly for 2010 and regressed with the increased impedance of the travel time in the new metro line extensions (Tomasiello et al., 2019). The network changes comprise the new metro and train stations opened after 2000 (Siqueira-Gay et al., 2017). The main differences between the two years are new stations in: (i) yellow metro line, (ii) lilac metro line, (iii) green metro lines and, (iv) train stations (Figure 2). For the accessibility indicators, the same urban service (e.g. hospitals and others) was used in accessibility measures for both years; therefore, the changes in accessibility levels reflect the changes in the transport network alone.

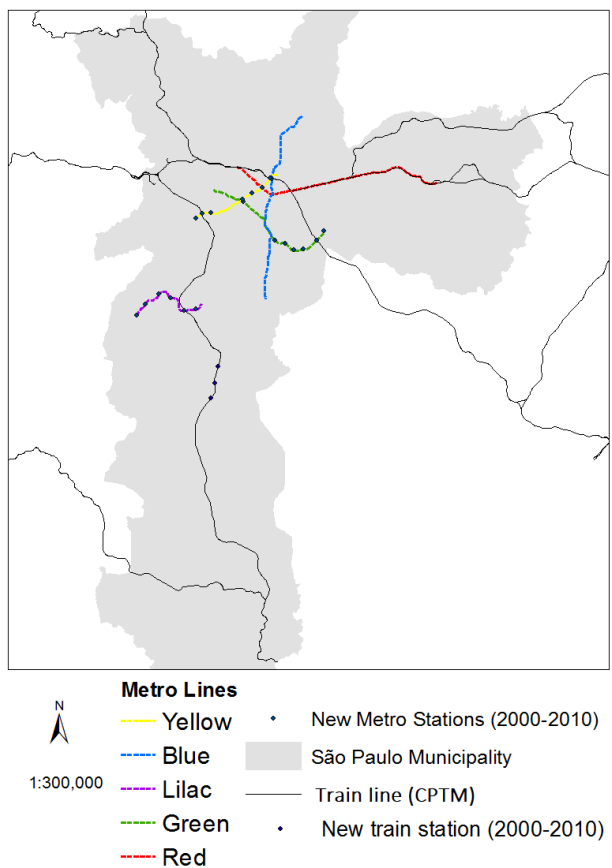

**Figure 2 – Metro and train transit lines in the São Paulo municipality: yellow and lilac lines were built after 2000**

The accessibility metric used was the cumulative opportunities to evaluate the potential number of opportunities to be reached, given a travel time (Neutens et al., 2010; Páez et al., 2012). The data used for calculating the indicators were: (i) the network with public transport travel time to calculate the travel time and to generate the service areas, (ii) the centroids of census tracts to be the center of services areas and (iii) the services facilities. First, the service area is generated based on a given travel time with the network and then, the number of urban services in each service area is counted. Then, the sum of the number of services to be reached is associated to the census polygon of their respective centroid (Figure 3).

The time over which people are willing to travel depends on the trip purpose. The threshold value is calculated based on the guidelines of the Department for Transport Business Plan (2012) from the UK and represents the median of all the trips using public transportation with specific purposes: education for accessibility to public and private schools; health for accessibility to hospitals and health centers and; leisure for accessibility to cultural facilities and sports centers. In 2000 and 2010, the values were similar, and the decision was to use the larger one to maintain the most conservative threshold. Figure 3 summarizes the steps for constructing the indicators and Table 1 presents the final measures. For the travel time calculation, one working day (Wednesday) at the peak hour (8 a.m.) is considered as reference.

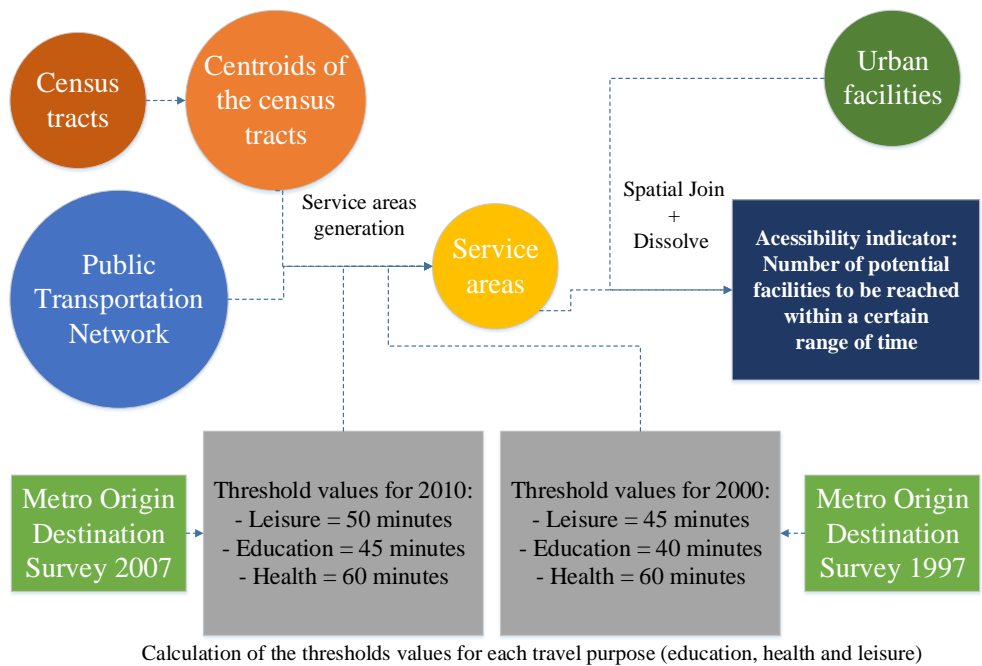

Figure 3 - Main steps of accessibility measures

Table 1 – Accessibility measures

| Urban services | Indicator |
|---|---|
| Hospitals | Number of hospitals to be reached within 60 minutes of travel time by transit |
| Health centers | Number of health centers to be reached within 60 minutes of travel time by transit |
| Public schools | Number of public schools to be reached within 45 minutes of travel time by transit |
| Private schools | Number of private schools to be reached within 45 minutes of travel time by transit |
| Sports centers | Number of sports centers to be reached within 50 minutes of travel time by transit |
| Culture facilities | Number of museums and libraries to be reached within 50 minutes of travel time by transit |

For capturing the social inequality, the analysis considers the value of average monthly income of the householder in minimum wages m.w. (in 2000 it was R$ 151,00[1] and in 2010, R$ 510,00[2]) from the census (Figure 4).

---

[1] In 2000, the reference value was 1 US$ = 1.08, resulting in the total amount of US$ 162.78
[2] In 2010, the reference value was 1 US$ = 1.95, resulting in the total amount of US$ 994.50

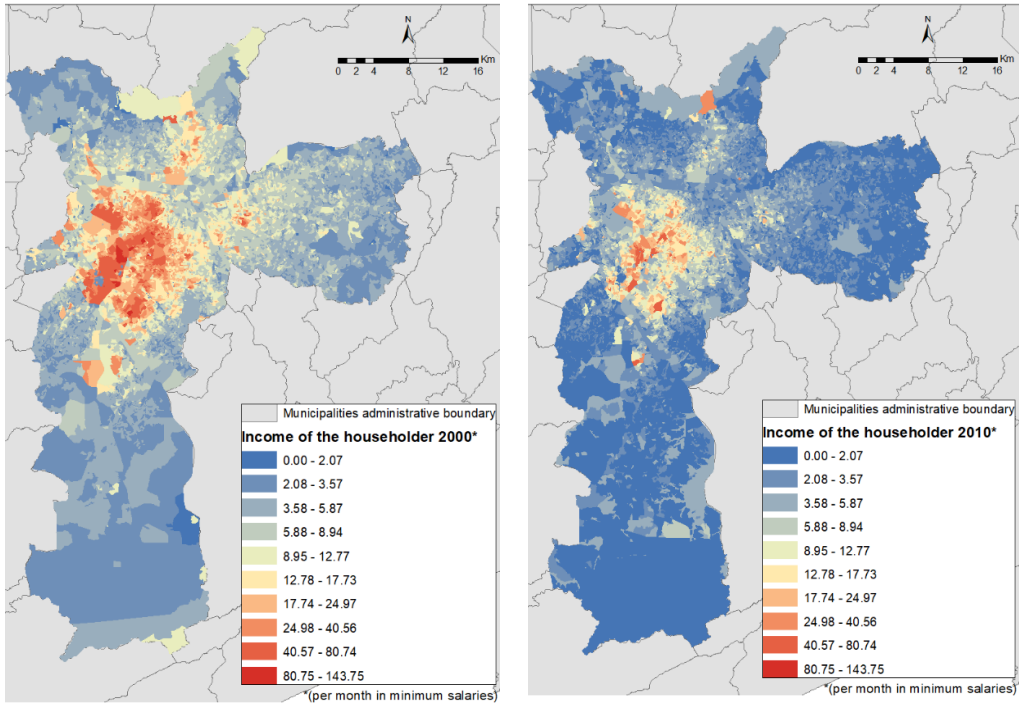
**Figure 4 – Spatial patterns and values of income in 2000 and 2010**

### 2.2. Dimensionality reduction

Firstly, dimensionality reduction was applied by using the Principal Component Analysis (PCA). PCA is a well-known technique for dimensionality reduction, which transforms the high dimensional data into a new one, which partially explains the variance of the original dataset. Based on the correlation matrix, the eigenvalues and eigenvectors are calculated to discover the main structure of the data. The first main direction explains the largest variability in the original dataset. The second, necessarily perpendicular to the first, explains the second largest variability and so on. All of them represent a transformed basis with every component perpendicular to each other (Zaki and Meira, 2013). The eigenvalues are the variance explained by each eigenvector. Therefore, the latter is the principal direction. Each principal component can be written as the linear decomposition of the original features and the weights or loads are representative of the feature importance in relation to each component. Relevant applications of PCA to dimensionality reduction were found in the social sciences, especially in the social vulnerability index formulation (Cutter et al., 2003). It is broadly used in the composite indicators analysis (Beccari, 2016) as well as in urban planning, as part of the procedure to classify city areas (Ibes, 2015). Especially for these studies, the reduction aims at removing the noise and correlated variables from the original dataset.

### 2.3. Groups composition

A cluster is defined as a group of similar instances. There are two types of distance measures, intra and inter clusters. The goal of the clustering procedure is to minimize the within-group similarity (intra cluster distance) and to maximize the distance between distinct clusters (inter cluster distance) (Joseph et al., 2016). They are useful to quantify how different the groups are and to compare the similarity or distance between each attribute. The dissimilarity between clusters is intrinsically related to the accuracy of the algorithm. The more distinct the groups are, the better the cluster assignment is.

One of the simplest algorithms for grouping instances is K-Means. Its advantage is the low computational cost and easiness of implementation. The application of clustering algorithm K-Means aimed at minimizing the distance between clusters. It involves two main steps: the cluster assignment and the centroid update. Firstly, the K number of clusters is set, and each point should be assigned to the closest mean. Therefore, the groups of points close to the mean value constitute one cluster. In the next step, the centroids of each cluster are updated. The convergence occurs if the cluster centroid does not change from one given interaction to the other. The algorithm has been enhanced since its publication in Macqueen (1967). Some applications of K-Means can be verified in the remote sensing area for image segmentation (Dhanachandra et al., 2015) and precipitation estimation (Mokdad and Haddad, 2017) as well as the improvements in the algorithm for time series analysis (Huang et al., 2016). This algorithm, associated with the previous dimensionality reduction, provides a comprehensive group analysis (Ibes, 2015).

## 3. INEQUALITIES ANALYSIS

The first step calculates the accessibility indicators and income. In the pre-processing procedure, the missing values of the income variable were replaced by the overall mean. In the 2000 database, there are 13278 census instances and in 2010, 18953. The difference in the number of census tracts is due to the changes in the urban area and population growth over the years. Because of these changes, the interviewer changes his/her reachable area for interviewing the population. The missing values represent less than 1% of all the data in 2000 and 3% in 2010. The descriptive statistics of the 2000 database are depicted in Table 2 and the 2010 one in Table 3. The accessibility levels to public and private schools on both years are different mainly due to the higher number of private schools. As the accessibility to culture facilities is highly concentrated in the city center in 2000 and 2010, the distribution encompasses a wide range of values and, therefore, a high standard deviation. The income indicators present the greatest difference between the years, with a higher average and median in 2000.

Table 2 – Descriptive statistic of 2000 dataset

|  | Access Hospitals | Access Sports Centers | Access Private Schools | Access Health Centers | Access Public Schools | Access Culture | Income |
|---|---|---|---|---|---|---|---|
| Average | 18.20 | 9.12 | 296.27 | 33.70 | 178.15 | 253.13 | 9.76 |
| Median | 12.00 | 7.00 | 261.00 | 32.00 | 167.00 | 74.00 | 6.05 |
| Standard deviation | 16.57 | 7.03 | 192.82 | 16.19 | 89.05 | 358.31 | 9.99 |
| Interval | 72.00 | 49.00 | 1581.00 | 139.00 | 789.00 | 1355.00 | 165.06 |
| Minimum | 0.00 | 0.00 | 0.00 | 0.00 | 1.00 | 0.00 | 1.62 |
| Maximum | 72.00 | 49.00 | 1581.00 | 139.00 | 790.00 | 1355.00 | 166.68 |

Table 3 – Descriptive statistic of 2010 dataset

| | Access Hospitals | Access Sports Centers | Access Private Schools | Access Health Centers | Access Public Schools | Access Culture | Income |
|---|---|---|---|---|---|---|---|
| Average | 17.26 | 8.28 | 277.64 | 31.87 | 167.30 | 243.85 | 4.24 |
| Median | 10.00 | 7.00 | 234.00 | 30.00 | 156.00 | 70.00 | 2.46 |
| Standard deviation | 16.66 | 6.89 | 196.27 | 16.50 | 90.17 | 354.13 | 5.02 |
| Interval | 72.00 | 49.00 | 1599.00 | 141.00 | 788.00 | 1359.00 | 143.75 |
| Minimum | 0.00 | 0.00 | 0.00 | 0.00 | 0.00 | 0.00 | 0.00 |
| Maximum | 72.00 | 49.00 | 1599.00 | 141.00 | 788.00 | 1359.00 | 143.75 |

The original dataset presented seven variables: six accessibility measures and the monthly income in m.w. The PCA performance resulted in 97% of the dataset being explained by 4 main components (Table 4). The first main component (C1) explains about 61% of the variance in the original dataset and presents lower loads to all variables comparing with the other component's loads. The first main component (C1) explains about 61% of the variance in the original dataset and presents lower loads absolute values to all variables comparing with the other component's loads (Figure 5). The map shows this component highly correlated to the urban fringe, following the income pattern (Figure 4) according to the highest load imputed to the income variable (Figure 6). The map reveals that income influences inversely the trends in the dataset, because as lower the income, higher is the principal component load (Figure 6). The second component (C2) explains 26% of the original variance, accumulating with the first the equivalent of 87% of the total. This component presents a strong dependence with the city centrality (Figure 6) and is highly positively correlated to high accessibility to hospitals, culture and high income (Figure 5), following the pattern distribution of the main metro stations in the São Paulo municipality (Figure 2). In other words, the first component can be understood as the dimension that explain the low-income distribution and the second, the high accessibility levels, both opposites that comprise the main variation of the total dataset. The other two components (C3 and C4) respectively present correlation to income, accessibility sports and health centers and public and private schools. The C3 map shows a distinguished pattern to the city center and C4 does not present a clear difference related to the city center and peripheral region.

Table 4 – Cumulative percentage of variance of the 2000 dataset

| Component | Cumulative percentage of variance explained by each component |
|---|---|
| C1 | 0.61 |
| C2 | 0.87 |
| C3 | 0.94 |
| C4 | 0.97 |

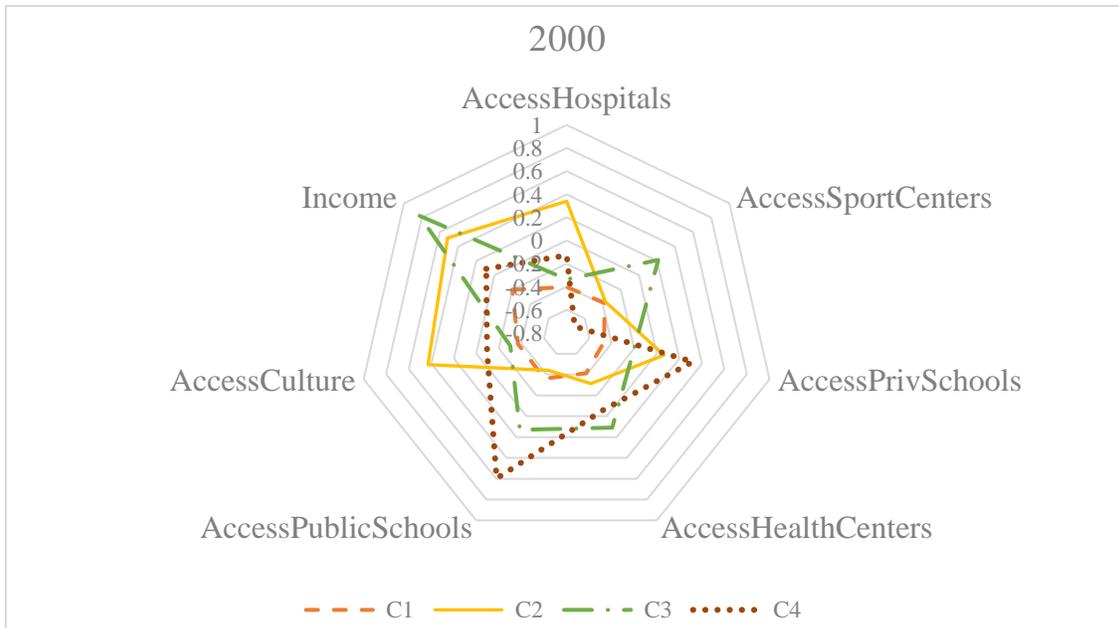

**Figure 5 – The proportion of the original variables in each eigenvector for the 2000 dataset**

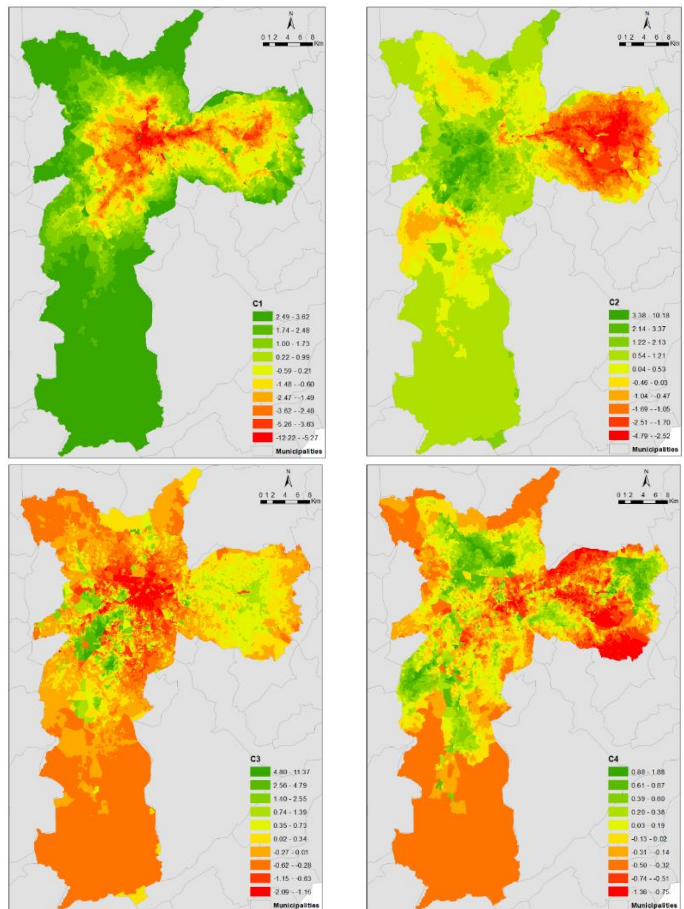

**Figure 6 – Principal components of the 2000 dataset**

The transformation applied to the 2010 dataset also reveals four main components explaining 96% of the original variance (Table 5). The first component (C1) also explains 61% and presents similar highest absolute load value to income (Figure 7) and spatial pattern (Figure 8) as the first component of the 2000 dataset. The same occurs with the second principal component (C2), which explains, jointly with the first, about 85% of the original variance. The main differences are in the fourth component (C4), which is also related to public and private schools. However, the map (Figure 8) does not show the same spatial patterns of 2000 (Figure 6), revealing the main discrepancies between 2000 and 2010 datasets.

Table 5 - Cumulative percentage of variance of the 2010 dataset

| Component | Cumulative percentage of variance explained by each component |
|---|---|
| C1 | 0.61 |
| C2 | 0.85 |
| C3 | 0.93 |
| C4 | 0.96 |

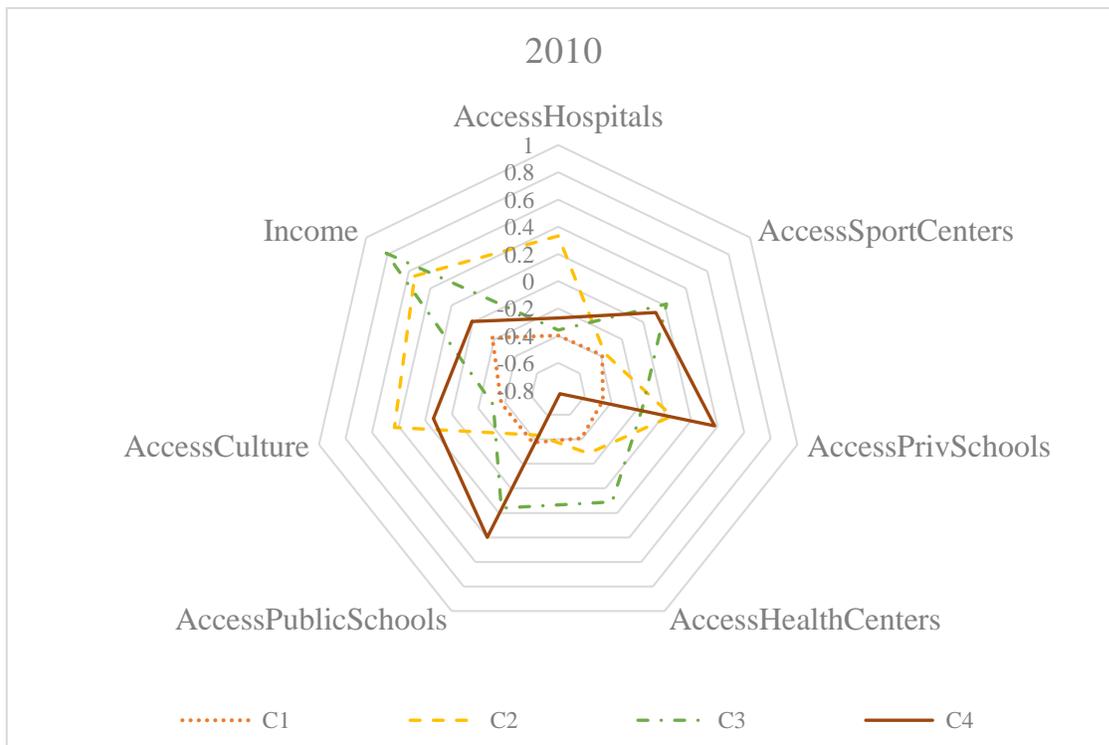

Figure 7 - The proportion of the original variables in each eigenvector for the 2010 dataset

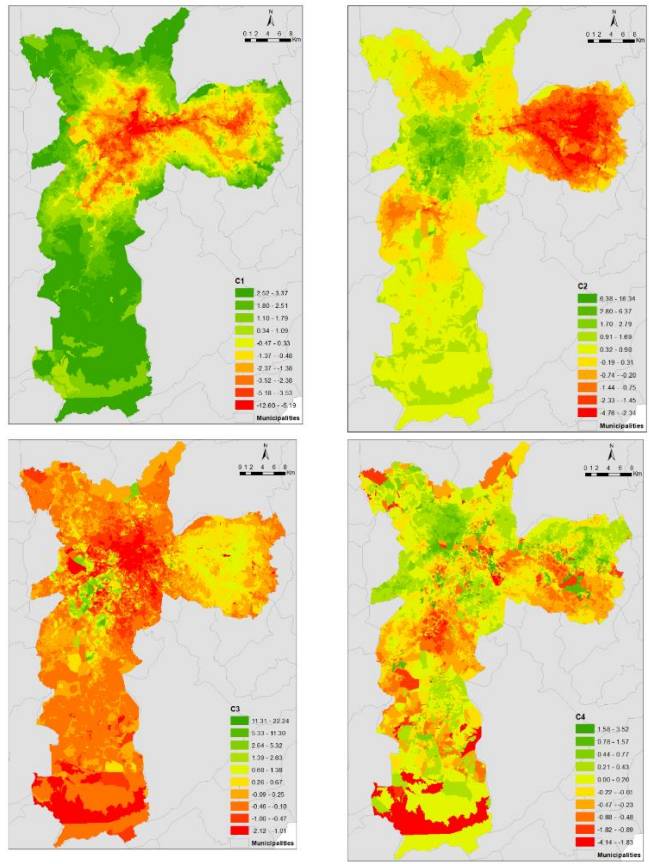
**Figure 8 – Principal Components of the 2010 dataset**

The two first main components, both with the cumulative percentage of variance of 87% in 2000 and 85% in 2010, were chosen to be the new dataset for clustering analysis. The similarity between the two components in the two years of analysis resulted in a similar number of clusters and groups patterns for 2000 and 2010. For the clustering analysis using K-Means, the best number of clusters was determined as seven for both years (Figure 9), number from which there is no considerable change in the sum of squared error. Table 6 depicts the results from the model parameters in clustering assignment. The initialization "random" method presents similar error as K-Means++ but a higher number of iterations. Considering the computational effort, the execution time is 0.24 seconds; therefore, there is no considerable difference between the two methods. In relation to the type of distance considered between the clusters, the Euclidean distance presents the smallest sum of squared errors in relation to the Manhattan distance implemented in Weka. The consideration of only two first main components instead of the original dataset also considerably decreases the sum of squared errors (from around 170 in 2000 and 121 in 2010 to around 40 in both years) and the time to build the model (0.5 to 0.24).

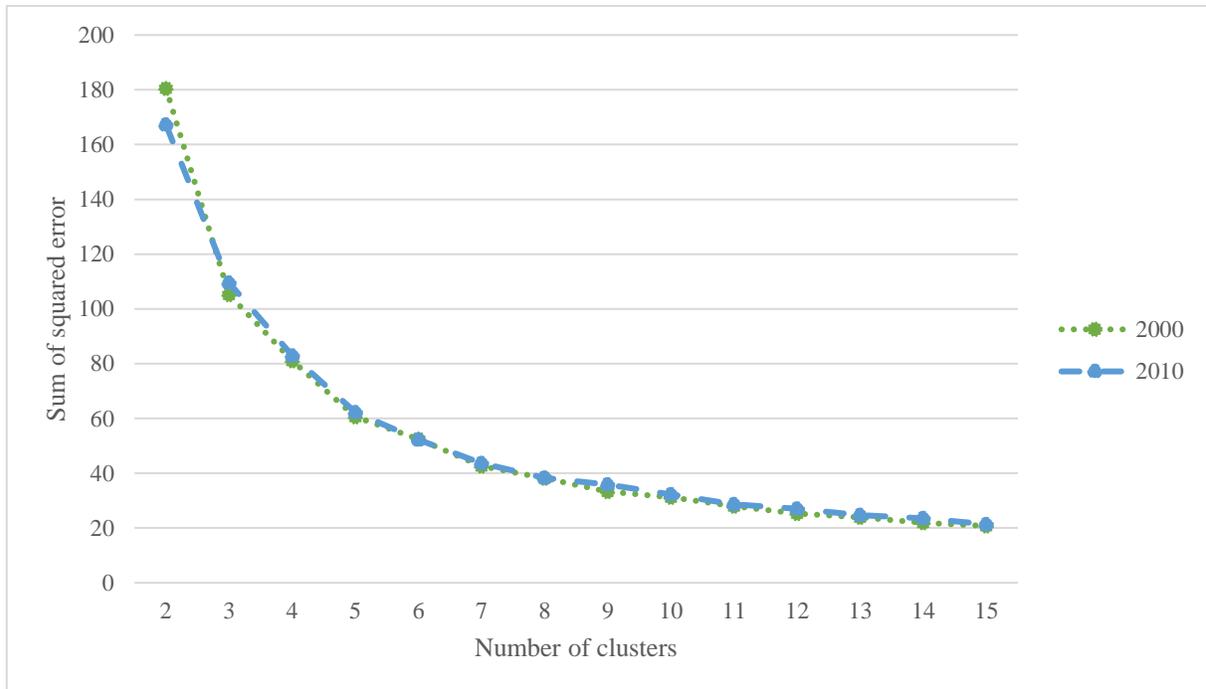

**Figure 9 – Elbow curve**

**Table 6 - Models parameters**

| K-Means | 2000 | 2010 |
|---|---|---|
| Distance Function | Euclidean | Euclidean |
| Initialization method | Random | Random |
| Number of clusters | 7 | 7 |
| Number of iterations | 47 | 55 |
| Within cluster sum of squared errors | 42.57 | 43.67 |

For composing the groups, they were divided into high and low-income according to the median value of each year. For 2000, the value of reference was 6.05 and for 2010, 2.46. In 2000, the low-income groups, generally present a low value of accessibility to hospitals and cultural facilities (Figure 10). Especially, low-income cluster number seven in the urban fringe has 27% of the São Paulo population and presents the lowest level of accessibility to all services. Conversely, the cluster that connects the city center and the east zone (number five), present a low-income level but it is very well located to access hospitals and cultural facilities. The total number of low income group corresponds to 80% of population and 22% of these (clusters number six and two) have better accessibility condition. Therefore, the good condition of accessibility and quality of life is accessed by a small proportion of population.

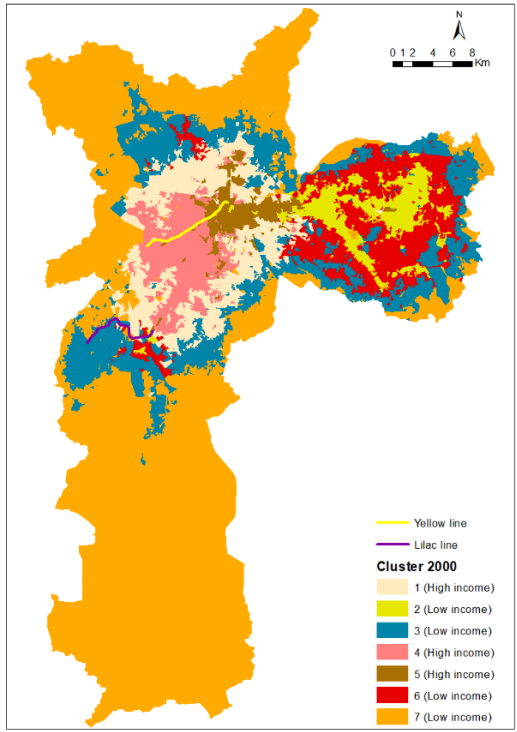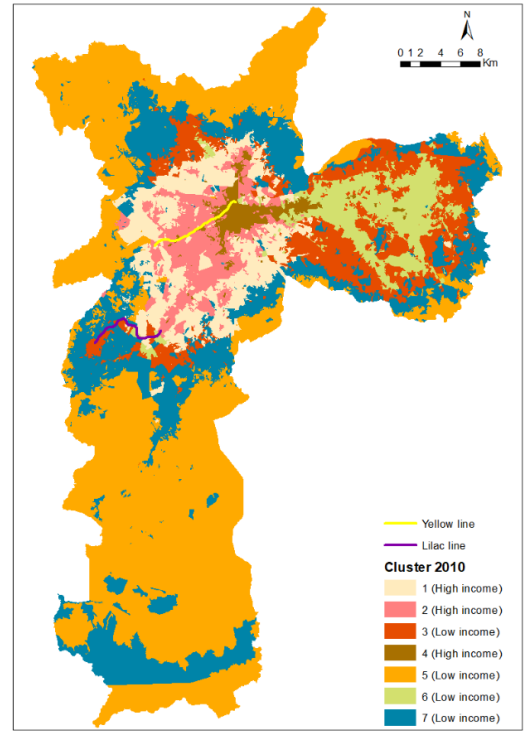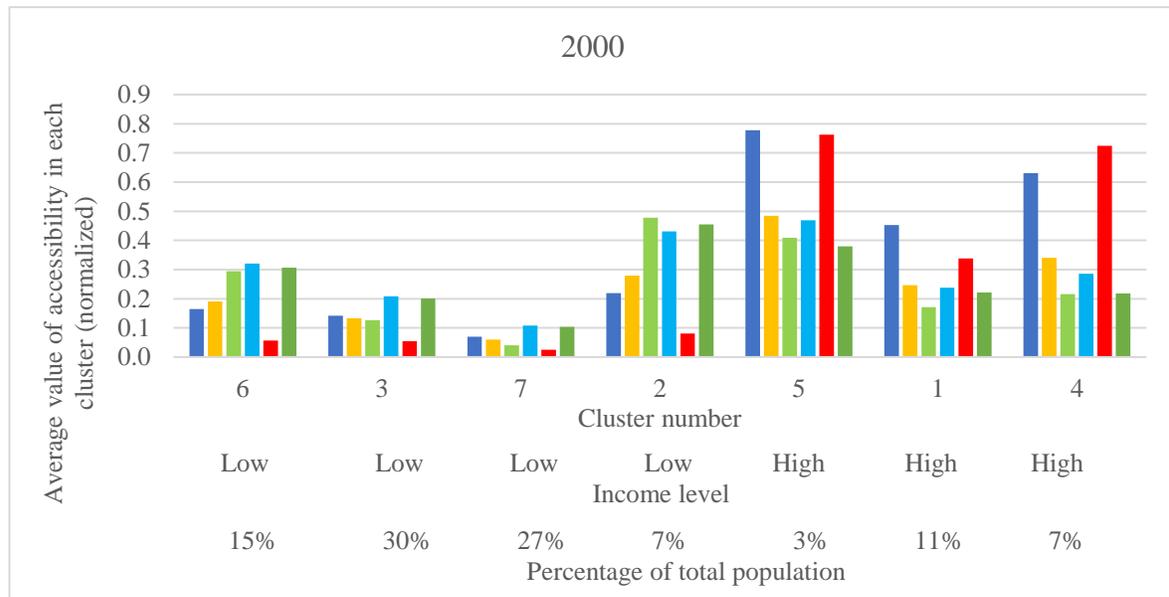

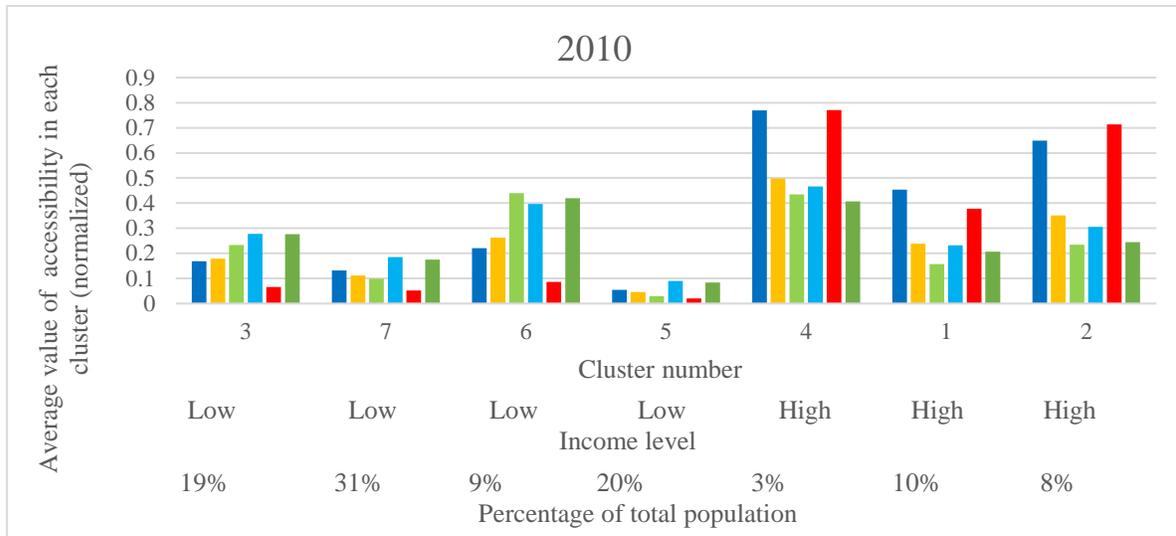

**Figure 10 – Groups composition in 2000 and 2010**

Considering the analysis of the 2010 groups, some of them present a similar spatial pattern to those of 2000 (they have the same color in Figure 10). For instance, cluster number one in both years represents the border of the city center. In 2000, cluster number seven presents a similar spatial pattern as number five in 2010 and both present the lowest level of accessibility to all services and low income. In both years, the general tendency in the low-income cluster is the low level of accessibility to cultural facilities and hospitals. Besides, the low-income groups represent 80% of the total population in 2010 and in 2000. In 2000, 27% of the population show the lowest level of accessibility to all services and this percentage decreases to 20% in 2010.

Especially in the surroundings of the new metro lines constructed between 2000 and 2010, the lilac line, in the south region of the São Paulo municipality (Figure 11), presents the predominance of low level of accessibility to all services. In 2010, it moves to the groups with low level of accessibility mainly to cultural facilities. In turn, in the yellow line, in the São Paulo center (Figure 10), a similar spatial pattern is verified in the groups' dispersion (cluster number four and five in 2000, and two and four in 2010) in both years, with groups of high income and high accessibility to all services. The provision of yellow line metro stations in the central area seems to reinforce the already consolidated pattern of radial configuration concentrating transport infrastructure in the city center, as accessibility level does not considerably change.

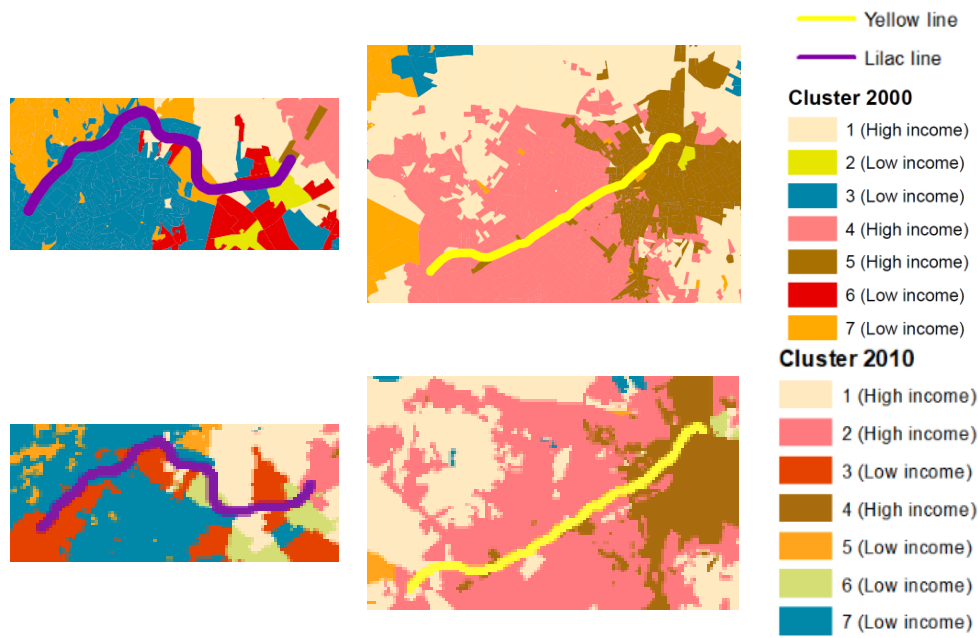

**Figure 11 – Surroundings of metro stations in 2000 (above) and 2010 (below)**

## 4. DISCUSSION

By using a multidimensional approach, the accessibility indicators and socioeconomic information were explored to understand the heterogeneity of opportunities and inequalities in the territory. In 2000, cluster number four in the city center with high accessibility, especially to hospitals and culture, follows the pattern described by Marques (2005) as a group with high income, high scholarly rate, best infrastructure, with elderly and a decreased growth rate. This centrality was also captured by Sposati; Monteiro (2017) as a group with high level of social exclusion.

Note that this centrality presents the smallest number of inhabitants (Figure 10); therefore, the good condition of accessibility and quality of life is accessed by a small proportion of the population. The analyses demonstrate that this region, which received the new stations of the yellow line, does not present the most unequal population group. However, the most deprived groups could use the line to access other regions of the city. Cluster number seven in 2000 and five in 2010 (Figure 10) present the lowest level of accessibility and, in the analysis of degree of exclusion presented by Sposati; Monteiro (2017), they comprise the districts with the most intense degree of social exclusion.

An interesting comparison between the result of Sposati; Monteiro (2017) and that obtained by the present work highlights the groups in the border of the city centrality - number one in 2000 and 2010 (Figure 10). They present moderate social exclusion and are classified with high income, high access to hospitals, culture and intermediate level of access to other services. The total of 10% of the population live in that region and are considered homogenous groups by Sposati; Monteiro (2017).

The groups in the east zone present a severe degree of social exclusion but considering the accessibility level, they do not present the worst condition (Figure 10). They are classified as low-income and represent about 20% of the city population (Figure 10) and live close to metro stations and bus corridors. Besides, the supply of some services, such as schools and health centers, is not critical. The quality of the service is not considered in the accessibility measures, but rather only the presence or absence of infrastructure.

Regarding the distribution of urban services, especially schools, Moreno-Monroy et al. (2017) highlight the inequalities related with income and accessibility to school in the Metropolitan Region of São Paulo and the importance of its being addressed in public policies. Especially considering the reality of the east zone of the city, Érnica; Batista (2012) investigate the relation between neighborhood and education quality. They state that the more vulnerable the surroundings of the school, the worse the quality of educational opportunities is. Therefore, even the east zone providing a considerable number of public and private schools, the people living in that region do not necessarily access a good educational service.

In São Paulo, it is possible to verify the educational inequalities throughout the city. For instance, the population with the highest income level have access to higher education and mostly attend private schools (Perosa et al., 2015). This high-income group living in the centrality is supplied with metro stations and present the highest values of private school accessibility. However, this supply is wasted by a group that does not need to reach different regions because they live close to the opportunities. The lilac line, in the south, is the opposite. The population living there need to reach the opportunities in the centrality and do not present a high level of accessibility to public schools. Further works should explore the differences between the groups considering their age and the daily differences of demand for educational purposes for a better diagnosis.

In our research, the most critical facility for the low-income population is the cultural centers. This result evidences the need for providing more cultural opportunities in the peripheral regions. A different behavior can be verified by the spatial pattern of accessibility to sport facilities. Higgs; Langford; Norman (2015) highlight the importance of measuring the proximity with sport facilities as a proxy of the quality of life and interaction. In the groups composition, the one in the east zone - number two in 2000 and number six in 2010 (Figure 10) – even classified with low income and in a vulnerable region of the city, present the highest level of accessibility to sports facilities.

Considering health opportunities, the accessibility to hospitals is highly correlated with accessibility to culture, therefore with high values concentrated in the city centrality (Figure 10). The accessibility to health centers is more widely spread across the city, in the east and south region (Figure 10). Hence, in the low-income cluster, this type of opportunity does not present a very critical accessibility level. Neutens (2015) presents a review of geographical accessibility to health services. He also presents the main challenges of assessing this type of service, once it depends on the affordability and costs of the health services, their quality, availability, accessibility and travel impedance between the patients and the service, as well as accommodation. In this sense, there is a lot to do for further exploring this type of accessibility given the groups condition (age and gender, for example, besides income) in transportation plans.

Given the results presented, it is possible to demonstrate the potential of exploring the cluster heterogeneity. The outcomes for planning practice are: (i) identify the heterogeneity of groups composition; (ii) identify the lack of urban services of deprived groups; (iii) analyze the consequences in the groups composition and level of services considering different alternatives of infrastructure provision and land use. Urban and environmental planning can be supported considering different instruments at: (i) the strategic level, such as Strategic Environmental Assessment (Siqueira-Gay and Sánchez, 2019); (ii) regional level, such as mobility and master plans (Malvestio et al., 2018); (iii) local level, such as Environmental Impact Assessment (Sánchez and Silva-Sánchez, 2008). Other sectoral initiatives, such as Transit Oriented Development, can be informed about hidden inequalities using public and available information.

The techniques, using or not the geographical information in the model, can be explored by planners as a powerful tool to acquire knowledge about the territory and its inequalities. Although this analysis presents potential information, some restrictions should also be considered, such as the Modified Area Unit Problem (MAUP). Recent research points out the need for cross scale effects evaluation, especially considering the land use-travel interaction (Kwan and Weber, 2008). Kwan; Weber (2008) highlight the invariance of space-time accessibility with the unit of analysis, therefore, the relations do not show substantial variation at different geographic scales. Pereira et al. (2018) demonstrate the dependence on an equity assessment to spatial scale and area unit. However, the authors highlight that the most suitable area identified can change according to the accessibility measure and opportunity analyzed. Therefore, there is no consensus in the literature and each region or case study should be analyzed in detail. For the planner's practice, this restriction should be considered, and the scales set according to the level of the project details.

## 5. CONCLUSIONS

This work aimed at identifying and describing patterns hidden in the distribution of accessibility to different urban services and socioeconomic information using a multidimensional approach. The analysis identified the most deprived groups in terms of income and accessibility. Although the low-income group present low accessibility level to hospitals and cultural facilities, an intermediate level of accessibility to sport centers and public schools in the São Paulo east zone reveals the heterogeneity in the city peripheries. By quantifying and mapping the complex distribution of transport, socioeconomic information and land use, urban plans can be informed about the general group's composition as well as identify lack of urban services of deprived groups. This methodological proposal can support future decisions for reducing inequalities and promoting the sustainable use of resources of environmental instruments at different levels of decision making as well as other sectoral initiatives. Future works can explore in more details how to integrate this information into policies and plans for systematically consider the environmental and social information and the equitable distribution of people and resources.

### ACKNOWLEDGEMENTS

This study was financed in part by the Coordenação de Aperfeiçoamento de Pessoal de Nível Superior - Brasil (CAPES) - Finance Code 001.